# Treatment and Aging Studies of GaAs(111)B Substrates for van der Waals Chalcogenide Film Growth

Mingyu Yu [1], Jiayang Wang [2], Sahani A. Iddawela [3], Molly McDonough [2], Jessica L. Thompson [3], Susan B. Sinnott [2,3,4,5,6], Danielle Reifsnyder Hickey [2,3,4], Stephanie Law [2,4,6,7,8,a]

[1]Department of Materials Science and Engineering, University of Delaware, 201 Dupont Hall, 127 The Green, Newark, Delaware 19716 USA
[2]Department of Materials Science and Engineering, The Pennsylvania State University, University Park, PA 16802 USA
[3]Department of Chemistry, The Pennsylvania State University, University Park, PA 16802 USA
[4]Materials Research Institute, The Pennsylvania State University, University Park, PA 16802 USA
[5]Institute for Computational and Data Science, The Pennsylvania State University, University Park, PA 16802 USA
[6]Penn State Institutes of Energy and the Environment, The Pennsylvania State University, University Park, PA 16802 USA
[7]2D Crystal Consortium Material Innovation Platform, The Pennsylvania State University, University Park, PA 16802 USA
[8]Department of Physics, The Pennsylvania State University, University Park, PA 16802 USA

[a] Electronic mail: sal6149@psu.edu

## I. INTRODUCTION

GaAs(111)B is a semiconductor substrate widely used in research and commercial fields due to its low cost, mature synthesis technology, and excellent properties for manufacturing electronic devices.[1-3] It is not only used to grow three-dimensional (3D) strongly-bonded materials,[4,5] but has also been used as a substrate for layered, van der Waals (vdW)-bonded chalcogenide film growth.[6-9] The GaAs (111)B substrate surface comprises a hexagonal lattice which matches the in-plane lattice symmetry of many vdW chalcogenide crystals such as GaSe and $MoSe_2$. Moreover, this 2D/3D heterostructure is conducive to leveraging the advantages of 2D and 3D semiconductors simultaneously in hybrid devices. One of the most common techniques for growing films on GaAs(111)B substrates is molecular beam epitaxy (MBE), which results in wafer-scale films with high purity, good crystallinity, and smooth surfaces. However, GaAs(111)B wafers cannot be directly used for growing epitaxial vdW chalcogenide films for two reasons: (1) the GaAs surface has a substantial number of dangling bonds that need to be passivated for vdW layers growth; (2) the substrate surface is covered with a thin epi-ready oxide layer which must be removed before film growth. Thermal deoxidation of GaAs substrates is typically performed at high temperatures under an As overpressure to compensate for the evaporation of As atoms from the substrate. However, group-V elements such as As are typically not available in chalcogenide MBE systems so as to minimize contamination. Therefore, investigating the thermal removal of the epi-ready oxide and the passivation of surface dangling bonds in a chalcogenide MBE system is needed to expand the use of GaAs(111)B substrates for the growth of vdW chalcogenide materials. Previous studies[8,10-12] have reported the use of a Se overpressure during thermal deoxidation of GaAs(111)B to prevent the formation of pits and Ga droplets as well as to terminate the surface with Se atoms, providing a passivated and deoxidized platform for growing epitaxial chalcogenide films such as CdSe[13,14] and ZnSe.[15]

In this paper, we optimize the method for deoxidizing GaAs(111)B substrates under a Se overpressure and successfully create a smooth, deoxidized, and passivated substrate for subsequent growth of vdW chalcogenide materials. The high reproducibility of this method has been verified via multiple trials. We also discuss the potential mechanism of Se-passivation on GaAs(111)B through first-principle calculations. Furthermore, we demonstrate the benefits of this method for the growth of vdW chalcogenide thin films using GaSe as a representative of vdW chalcogenides. In addition to deoxidation and passivation, aging of GaAs(111)B substrates is another concern. We find that severely aged substrates have difficulty maintaining a smooth surface during the deoxidation and passivation process and cause GaSe crystals to nucleate in random shapes and orientations. Food-grade vacuum packaging is found not to completely prevent this aging process. We describe a method using water droplet testing to determine the age of the substrate. Finally, X-ray photoelectron spectroscopy (XPS) characterization reveals that the natural aging of GaAs(111)B in the air results in an increase in surface oxides, $Ga_2O_3$ and $As_2O_3$, while exposure to ultraviolet (UV)-ozone not only enhances the contents of these two oxides but also generates a new oxide, $As_2O_5$. Our research contributes to expanding the compatibility of GaAs(111)B with diverse growth materials and the production of high-quality heterostructure devices.

## II. EXPERIMENTAL DETAILS

### A. Treatment of GaAs(111)B substrate & GaSe growth

We used 2" epi-ready GaAs(111)B wafers from WaferTech which were diced into 1 cm × 1 cm pieces. Each piece was degreased by sequential sonication in acetone, isopropanol (IPA), and de-ionized (DI) water for 10 min at room temperature. Immediately after cleaning, the wafer was loaded into the load lock chamber of a DCA R450 chalcogenide MBE system and degassed at 200 °C in $5 \times 10^{-7}$ Torr for 2 hours to eliminate any residual contaminants. We then transferred the wafer to the growth chamber for deoxidation, where we heated and annealed the wafer under a Se overpressure and then cooled it down. The heating/cooling rate was maintained at 30 °C min$^{-1}$, and the Se flux was always supplied when the substrate temperature was above 300 °C in order to suppress the substrate decomposition and formation of Ga droplets at high temperatures. Specific annealing temperatures, times, and Se fluxes will be discussed in Section IV-A. For comparison, the substrate of Sample #7 was deoxidized at 610 °C under an As overpressure of $1.12 \times 10^{-6}$ Torr for 10 min in a Veeco GENxplor III-V MBE reactor. The heating/cooling rate was maintained at 10 °C min$^{-1}$, and the As flux was supplied when the substrate temperature was above 300 °C. The deoxidized substrate was promptly transferred to the chalcogenide MBE system using a $N_2$-purged glove bag. We then grew GaSe films with a thickness of ~16.8 nm on differently deoxidized substrates using the same growth conditions. Ga and Se fluxes were provided independently from separate Knudsen effusion cells and were calibrated using a quartz crystal microbalance at the substrate position. The substrate temperature was measured by a thermocouple mounted behind the substrate. During thermal treatment of the substrate, *in-situ* reflection high energy electron diffraction (RHEED) was employed to monitor and confirm the removal of the substrate oxide layer. After growth, the samples were immediately sent for characterization. A subset of samples were treated with UV-ozone using a Boekel Scientific UV cleaner, as described in detail in Section IV-D.

### *B. Ex-situ characterization*

High resolution X-ray diffraction (HRXRD) 2θ/ω and ω scans were performed on a Malvern PANalytical 4-Circle X'Pert 3 diffractometer equipped with a Cu-Kα$_1$ source. 2θ/ω scans were used to identify sample phases, while ω scans offered insight into crystal defects. Sample surface morphology was observed using a Bruker Dimension Icon AFM. To study the effect of substrate aging on crystalline film growth, electron-transparent cross sections were extracted using an FEI Scios 2 dual-beam focused ion beam (FIB). The cross sections were analyzed via annular dark-field scanning transmission electron microscopy (ADF-STEM) in a dual spherical aberration-corrected FEI Titan³ G2 60-300 STEM operating at 300 kV, with a probe convergence angle of 21.3 mrad and collection angles of 42-244 mrad. Surface composition analysis was performed using XPS that was measured on a Physical Electronics VersaProbe III instrument equipped with a monochromatic Al Kα X-ray source ($hv$ = 1486.6 eV) and a concentric hemispherical analyzer. The binding energy axis was calibrated using sputter cleaned Cu (Cu $2p_{3/2}$ = 932.62 eV, Cu $3p_{3/2}$ = 75.1 eV) and Au foils (Au $4f_{7/2}$ = 83.96 eV).[16] Measurements were made at a takeoff angle of 30 ° with respect to the sample surface, resulting in a typical sample depth of 2–4 nm. Quantification was conducted using instrumental relative sensitivity factors that account for the X-ray cross section and inelastic mean free path of the electrons. The analysis size was about 200 µm in diameter. Ion sputtering used 2 kV Ar$^+$ rasterized over a 2 mm × 2 mm area with a rate of 5 nm min$^{-1}$.

## III. COMPUTATIONAL DETAILS

First-principal calculations of Se-passivated GaAs(111)B surface models were performed using the Vienna ab initio simulation package,[17] which implements density functional theory and a plane-wave basis set with the projector-augmented wave method.[18] The exchange-correlation functions were approximated through generalized gradient approximation as stated in Perdew-Burke-Ernzerhof parametrization.[19] The valence electron configurations are $4s^24p^1$ for Ga, $4s^24p^3$ for As, and $4s^24p^4$ for Se. The plane wave cutoff energy was set to 600 eV and the Monkhorst-Pack k-mesh was sampled with a density of 0.05 Å s$^{-1}$. As for structure relaxation, the thresholds for determination of convergence were using 10$^{-5}$ eV as energy break conditions for the electronic self-consistence loop and Hellmann-Feynman force on each atom is less than 0.01 eV Å$^{-1}$. To describe surface geometry, the 10-atomic-layer slab model was generated with 18 Å thickness of vacuum space.

## IV. RESULTS AND DISCUSSION

### A. *Optimization of thermal treatment of GaAs(111)B in Se*

For the thermal deoxidation of GaAs(111)B under a Se overpressure, three parameters may affect the resultant surface quality: annealing temperature, time, and Se flux. The ideal conditions would produce a substrate surface that is completely deoxidized and as smooth as possible. Table 1 summarizes the processing conditions for six GaAs(111)B samples thermally deoxidized under a Se flux.

TABLE 1. Deoxidation parameters for GaAs(111)B substrates using a Se flux.

| Sample | Parameters | | |
|---|---|---|---|
| | Se flux [×10$^{14}$ atoms cm$^{-2}$ s$^{-1}$] | Annealing temperature [°C] | Annealing time [min] |
| #1 | 1.0 | 630 | 7 |
| #2 | 1.0 | 680 | 7 |
| #3 | 1.0 | 700 | 7 |
| #4 | 1.5 | 680 | 7 |
| #5 | 0.4 | 680 | 7 |
| #6 | 1.0 | 680 | 14 |

Due to the coverage of the thin oxide layer, the freshly loaded GaAs(111)B exhibited a blurry dashed RHEED pattern, as shown in Fig. 1(a). We first studied the annealing temperature by fixing the Se flux at $1 \times 10^{14}$ atoms cm$^{-2}$ s$^{-1}$ and the annealing time at 7 min. When the substrate temperature gradually increased to 630 °C, the RHEED pattern of Sample #1 became significantly clearer, and after staying at this temperature for 7 min, the dashed lines became more continuous, as shown in Fig. 1(b). However, the lines are still not entirely continuous, indicating that the temperature of 630 °C is not high enough to completely remove the oxide. The elevated annealing temperature of Sample #2 to 680 °C led to the sharp and streaky RHEED pattern shown in Fig. 1(c), which is typical of GaAs(111)B[20,21] and indicates a complete removal of the oxide layer. The resulting substrate surface was smooth without obvious defects, as shown in the AFM image in Fig. 2(a). Further raising the annealing temperature to 700 °C for Sample #3 resulted in significant evaporation of As atoms, creating numerous defects on the substrate surface, as shown in Fig. 2(b). Therefore, we determine 680 °C to be the ideal deoxidation temperature. Next, we studied higher and lower Se fluxes as well as longer annealing times at 680 °C. Fig. 2(c) of Sample #4 shows that an excess Se flux is not problematic as it does not persist on the surface at high temperatures. However, insufficient Se flux fails to adequately compensate for the loss of As during annealing, leading to surface defects as shown in Fig. 2(d) of Sample #5. Finally, prolonged annealing also causes a severe evaporation of As atoms, resulting in pits and Ga droplets, as shown in Fig. 2(e) of Sample #6 (the droplet features are more visible in the inset). Combining the RHEED and atomic force microscopy (AFM) results, we determined the optimal parameters to thermally deoxidize GaAs(111)B in a Se flux to be: Se flux $\geq 1 \times 10^{14}$ atoms cm$^{-2}$ s$^{-1}$, annealing temperature of 680 °C, and annealing time of 7 min. The optimal set of conditions produces fully deoxidized GaAs(111)B substrates with surface root mean roughness (RMS) as low as 0.47 nm and high reproducibility.

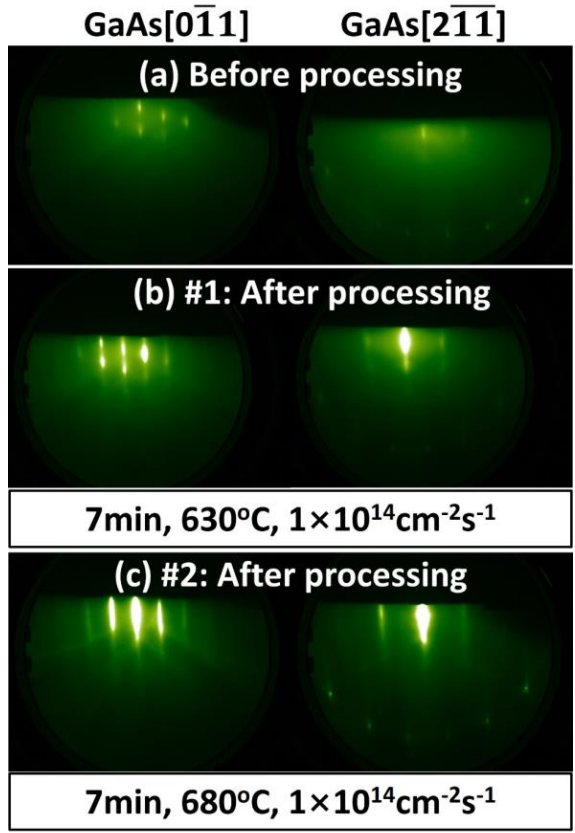

FIG. 1. RHEED patterns of (a) a freshly loaded GaAs(111)B substrate, (b) Sample #1, and (c) Sample #2 with the azimuthal incident electron beam along the $[01\bar{1}]$ (left column) and $[2\bar{1}\bar{1}]$ (right column) directions. Sample #1 and #2 are the GaAs(111)B substrates after being annealed in a Se flux of $1 \times 10^{14}$ atoms cm$^{-2}$ s$^{-1}$ for 7 min at 630 °C and 680 °C, respectively.

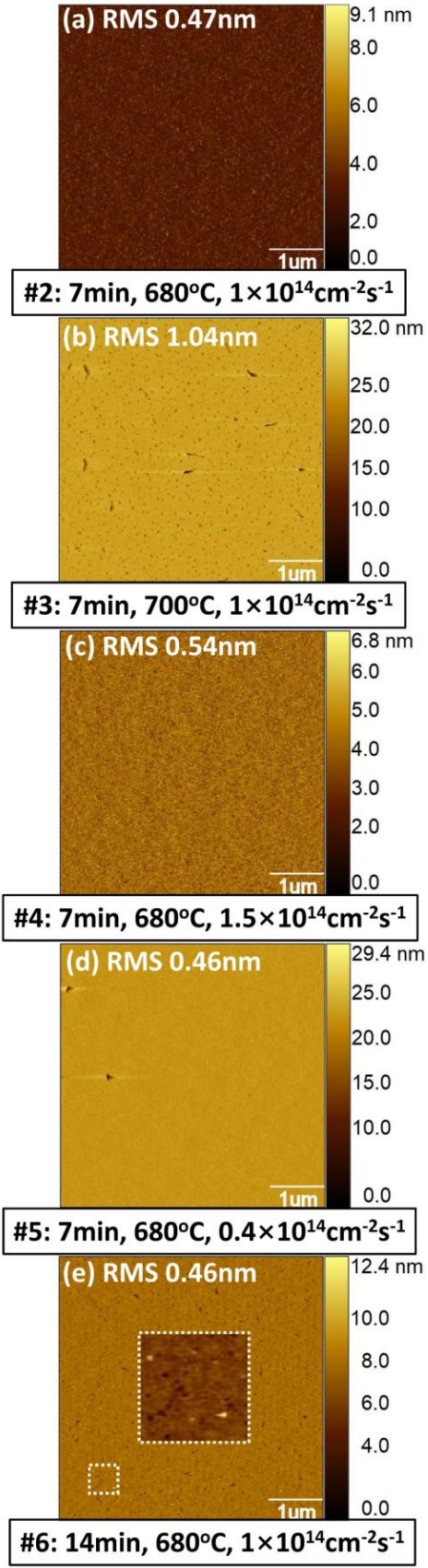

FIG. 2. AFM images of GaAs(111)B Sample (a) #2, (b) #3, (c) #4, (d) #5, and (e) #6 after being thermally treated in Se using different parameters, which can be seen in Table 1. The inset in (e) is a zoomed-in view of the white dashed box in the lower left corner showing the pits and droplet features more clearly.

## B. Theoretical model of Se-passivated GaAs(111)B surface

Numerous studies[8,10-12] have claimed that a benefit of deoxidizing GaAs(111)B substrates under a Se flux is that it simultaneously removes the oxide layer and passivates the top layer of the substrate with Se, which is useful for the subsequent growth of vdW chalcogenide films. Here we offer more comprehensive insights into the Se-passivation mechanism through first-principle calculations. Fig. 3(a) demonstrates the relaxed crystal structure of GaAs with an As-terminated top layer. The lattice constant of GaAs is calculated to be 5.76 Å, consistent with the theoretical value in the literature.[22] The As-terminated surface undergoes reconstruction compared to its bulk counterpart, resulting in a bond length between the surface As atom and the nearest Ga atom of 2.52 Å, which is 1.95 % longer than that in bulk GaAs. We consider two possible schemes for Se-passivation of the As-terminated GaAs(111) surface: (1) Se atoms are directly adsorbed onto the substrate surface; (2) Se atoms substitute the top As atoms and bond with the nearest Ga atoms. Both schemes are considered under the same series level of coverage. The stability of each configuration is evaluated by the heat of formation $H_f$, which is calculated by Eq. (1):

$$H_f = \frac{E_{slab} - n_{Ga}E_{Ga_8}^{bulk} - n_{As}E_{As_8}^{bulk} - n_{Se}E_{Se_{32}}^{bulk}}{n_{total}} \quad (1)$$

where $E_{slab}$ is total energy of doped systems; $E_{Ga_8}^{bulk}$, $E_{As_8}^{bulk}$, and $E_{Se_{32}}^{bulk}$ are the chemical potentials of each atomic species under their most stable form; $n_{Ga}$, $n_{As}$, are $n_{Se}$ represent the number of Ga, As, and Se atoms in the supercell, respectively, and $n_{total}$ is the total number of atoms.

In the adsorption scheme, we first consider the energetically most favorable sites for individual Se atoms. Fig. 3(b) depicts three possible adsorption sites: hollow 1, hollow 2, and top. "Top" is directly above the As atoms, while "hollow 1" (Fig. 3(c)) and "hollow 2" (Fig. 3(d)) refer to the two positions in the groove composed of As atomic layers. The $H_f$ of Se adsorption at the three sites are 0.0300, 0.0210, -0.0008 eV atom$^{-1}$, respectively, indicating that the "top" is the most stable adsorption site. This could be attributed to the directionality of the lone pair electrons provided by As atoms. Therefore, we adopted the "top" site for the adsorption scheme. Then we compare the $H_f$ in the adsorption scheme and the substitution scheme under series of coverage. The selection of As atomic sites in both schemes was assumed to be random. The comparison results and the corresponding crystal structure models are shown in Fig. 4. The $H_f$ reaches a minimum when Se atoms replace 75 % of the surface As atoms, indicating that the surface composed of 75 % Se and 25 % As atoms obtained by the substitution scheme should have the most energetically stable state, which is consistent with previous reports.[11]

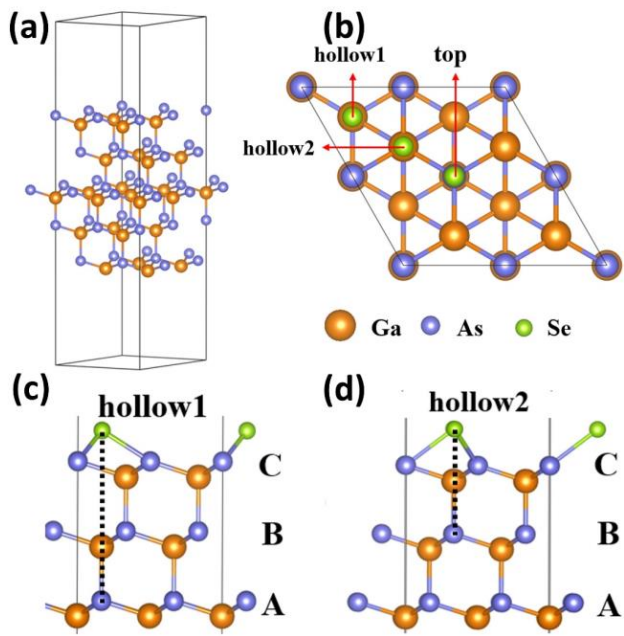

FIG. 3. (a) Crystal structure of GaAs slab model consisting of 10 atomic layers. (b) Top view of the GaAs(111)B surface lattice, where the three green balls represent the three Se adsorption sites on the As-terminated GaAs(111) surface: the "top" site is directly above the As atoms; the "hollow 1" and "hollow 2" sites are located at two different positions in the groove composed of As atomic layers. Side view of the GaAs(111)B crystal structure with Se adsorbed on the (c) "hollow 1" and (d) "hollow 2" sites. "A", "B", and "C" respectively represent three As sublattice layers that cannot completely overlap.

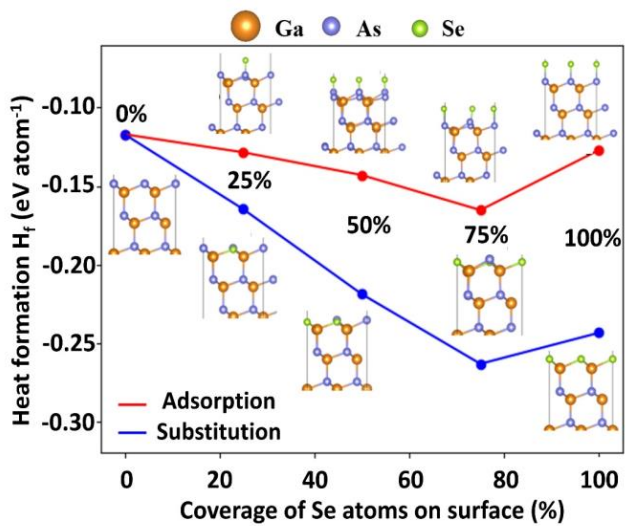

FIG. 4. Energy profile of two schemes for Se passivation on GaAs(111)B surface under different levels of Se coverage.

## C. *Importance of Se-passivation for vdW chalcogenide film growth*

Next, we demonstrate the importance of Se-passivation on GaAs(111)B substrates for the successful growth of vdW chalcogenide films. As a test case, we will discuss the MBE growth of GaSe, a typical vdW chalcogenide material. Since the GaSe crystal has a non-negligible lattice mismatch of ~ 6.4 % with GaAs(111),[23] forming crystalline GaSe films on GaAs(111)B substrates requires passivation of the substrate dangling bonds. We used the same conditions to grow GaSe films on two different substrates. The substrate of Sample #7 was deoxidized under As without Se passivation, while the substrate of Sample #8 was deoxidized under Se and passivated. The streaky RHEED pattern in Fig. 5(a) confirms that the substrate of Sample #7 is completely deoxidized under As. After depositing GaSe on both deoxidized substrates for 40 min at a rate of 0.07 Å s$^{-1}$, we found that GaSe growth failed on the As-deoxidized substrate; the 2θ/ω scan for Sample #7 (Fig. 5(b)) only shows two peaks belonging to the GaAs substrate. In comparison, the 2θ/ω scan for Sample #8 detects three additional peaks for the GaSe (002), (004), and (00$\underline{10}$) planes, respectively, confirming the formation of GaSe crystals. This experiment clearly shows that the simultaneous deoxidation and Se- passivation of GaAs(111)B substrates promotes the growth of vdW chalcogenide thin films.

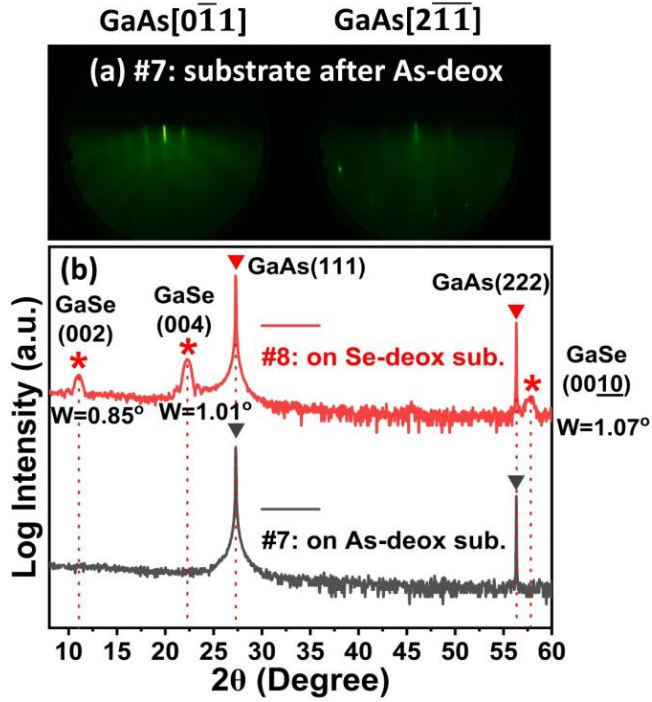

FIG. 5. (a) RHEED pattern of the As-deoxidized GaAs(111)B substrate (for Sample #7) taken along the $[01\bar{1}]$ (left column) and $[2\bar{1}\bar{1}]$ (right column) directions. (b) HRXRD 2θ/ω scans of GaSe Sample #7 and #8, whose substrates were deoxidized under As and Se, respectively. "∗" and "▼" mark the GaSe and GaAs peaks, respectively. "W" represents the value of the full width at half maximum (FWHM).

## D. Investigation on GaAs(111)B aging

Though it is common knowledge that GaAs(111)B is susceptible to aging in the air, its instability led to two unexpected findings during the course of this experiment: (1) even with optimized deoxidation/passivation treatment, the state of the GaAs(111)B substrate continues to affect the GaSe growth; (2) storing GaAs(111)B wafers in food-grade vacuum packaging does not prevent surface degradation. We will now discuss these two issues in detail.

Typically, the surface of GaAs(111)B wafers is covered by a thin epi-ready oxide layer. To avoid contamination, after dicing the wafer, we stored individual chips in polymer sample boxes, then vacuum-sealed these boxes using food-grade bags and placed them in a 5 °C refrigerator. Nevertheless, significant variations in the GaSe films grown on aged GaAs were observed. Fig. 6(a)–(c) illustrates the surface morphology of GaSe Sample #9, #10, and #11 deposited under the same conditions on substrates of different ages. Here, "fresh", "semi-aged", and "aged" refer to the duration from unpacking to usage, which is 7 days, 45 days, and 8 months, respectively. As the substrate aging progresses, the GaSe thin films became increasingly rough, accompanied by worse coalescence and more irregular-shaped nucleation. The change in the GaAs(111)B surface properties as a function of age was reflected in water droplet tests, as shown in Fig. 7(a)–(c), where the contact angle between the water droplet and the substrate surface noticeably increased with substrate aging, signifying a surface with higher hydrophobicity. We also confirmed that air exposure expedites this change. For instance, a 2-day air exposure can yield results comparable to those achieved by storing the substrates in vacuum-sealed bags for 8 months. More interestingly, UV-ozone cleaning has been found to restore surface hydrophilicity, as exemplified in Fig. 7(d).

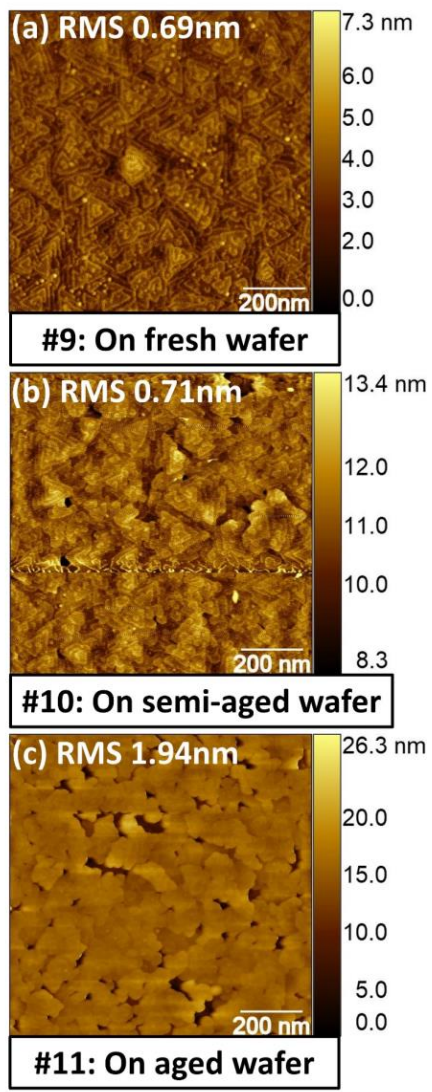

FIG. 6. AFM images of GaSe Sample (a) #9, (b) #10, and (c) #11. They were grown under identical conditions using substrates that were vacuum sealed in food-grade bags for 7 days ("fresh"), 45 days ("semi-aged"), and 8 months ("aged"), respectively. The height scale in (b) is adjusted to start from 8.3 nm instead of 0 for a clear visualization of the morphology.

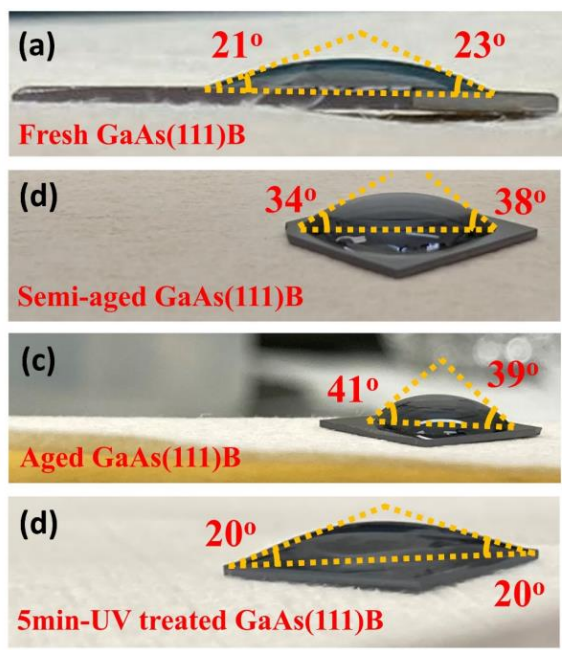

FIG. 7. Water droplet tests on GaAs(111)B substrates that is (a) fresh, (b) semi-aged, (c) aged, and (d) treated by UV-ozone for 5 min. The 5-min UV-ozone treatment was conducted on an aged substrate.

To understand the nature of the changes occurring on the substrate surface and their impact on hydrophobicity/hydrophilicity, as well as to understand why UV-ozone exposure restores hydrophilicity, XPS analysis was conducted on fresh, aged, and UV-treated GaAs(111)B substrates. The XPS spectra in Fig. 8 show that the aged substrate exhibits an increase of 48.6 % in $As_2O_3$ and 59.2 % in $Ga_2O_3$ compared to the fresh substrate. The 5-min UV-ozone treatment not only reduced organics by 55.7 % but also increased $As_2O_3$ by 92% and $Ga_2O_3$ by 47 %, and caused the formation of a new oxide, $As_2O_5$. Exposing the surface to UV-ozone for another 5 min further increased oxides and decreased organics. It is worth noting that extending the UV-exposure time increases the proportion of $As_2O_5$ in $As_xO_y$. The compositions of C, O, As, Ga, $Ga_xO_y$, and $As_xO_y$ are summarized in Table 2.

Since no additional findings were detected on the aged substrates other than additional oxides, and as exposure to air has been verified to accelerate the process, it is speculated that the aging is primarily caused by additional oxidation. The reason vacuum packing is unable to fully prevent oxidation is that food-grade vacuum packaging bags cannot achieve a high level of vacuum, and the residual air allows the aging process to continue slowly. As the oxide film composed of $As_2O_3$ and $Ga_2O_3$ becomes thicker and denser, the hydrophobicity of the GaAs(111)B surface increases.[24] As for UV-ozone treatment, it is well known that the cleaning mechanism involves generating highly active oxygen or ozone atoms to attack organic contaminants, converting them into volatile byproducts for removal, inevitably leading to further oxidation of the sample. The UV-treated GaAs(111)B surface here displayed two features: fewer organic molecules and the emergency of $As_2O_5$. While the reduction in organic impurities may contribute to the restoration of hydrophilicity, the fact that a sequential ultrasonic cleaning in acetone/IPA/DI water failed to produce the same effect suggests that the presence of $As_2O_5$ is a more credible explanation for the improved surface hydrophilicity.

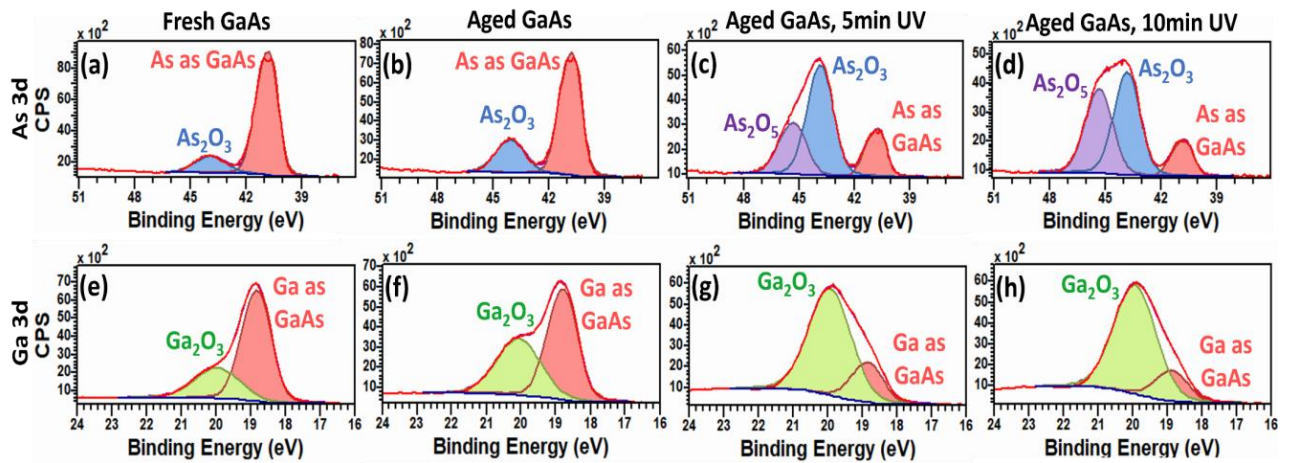

FIG. 8. XPS spectra of GaAs(111)B surface. (a-d) As 3d regions and (e-h) Ga 3d regions of (a, e) a fresh substrate, (b, f) an aged substrate, (c, g) an aged substrate after being treated by UV-ozone for 5 min, and (d, h) an aged wafer after being treated by UV-ozone for 10 min.

TABLE 2. Relative composition in atomic percent (%) obtained from the XPS spectra.

| GaAs(111)B | Composition | | | | | | |
|---|---|---|---|---|---|---|---|
| | C | O | As as GaAs | $As_2O_3$ | $As_2O_5$ | Ga as GaAs | $Ga_2O_3$ |
| Fresh | 28.0 | 27.5 | 17.4 | 3.5 | -- | 16.4 | 7.1 |
| Aged | 19.4 | 35.2 | 14.2 | 5.2 | -- | 14.8 | 11.3 |
| Aged, 5 min UV | 8.6 | 51.9 | 3.4 | 10.0 | 4.9 | 4.6 | 16.6 |
| Aged, 10 min UV | 8.1 | 53.9 | 2.3 | 8.1 | 6.9 | 3.7 | 17.1 |

To address the last question regarding how the substrate aging affects subsequent crystal growth under the condition of deoxidation/Se-passivation treatment, ω scans and STEM tests were performed on GaSe Sample #9 – #11. The broadening of the rocking curve peaks in Fig. 9(a) as the substrate aging progresses suggests that aging introduces more defects into the GaSe crystals, which has been further confirmed by the STEM images. Fig. 9(b) shows that the GaSe layers on a fresh substrate (Sample #9) have an ordered layer-by-layer epitaxial structure, while Fig. 9(c) reveals a substantial number of stacking faults and chaotic crystallographic arrangements within the GaSe film on the aged wafer (Sample #11), and the interface exhibits a coarse texture with numerous defects. We suspect that aging results in more oxides ($As_2O_3$ and $Ga_2O_3$) on GaAs(111)B surface, leading to significant damage to the surface during the deoxidation process. The substrate, characterized by more defects, promotes the formation of more stacking faults and misalignments within the GaSe layers. Finally, although UV-ozone cleaning effectively reinstates surface hydrophilicity, it does not yield an ideal platform for further growth. This is because the post-UV-treated surface accumulates more oxides, and the process of complete oxide removal causes severe damage to the surface, as depicted in Fig. 10.

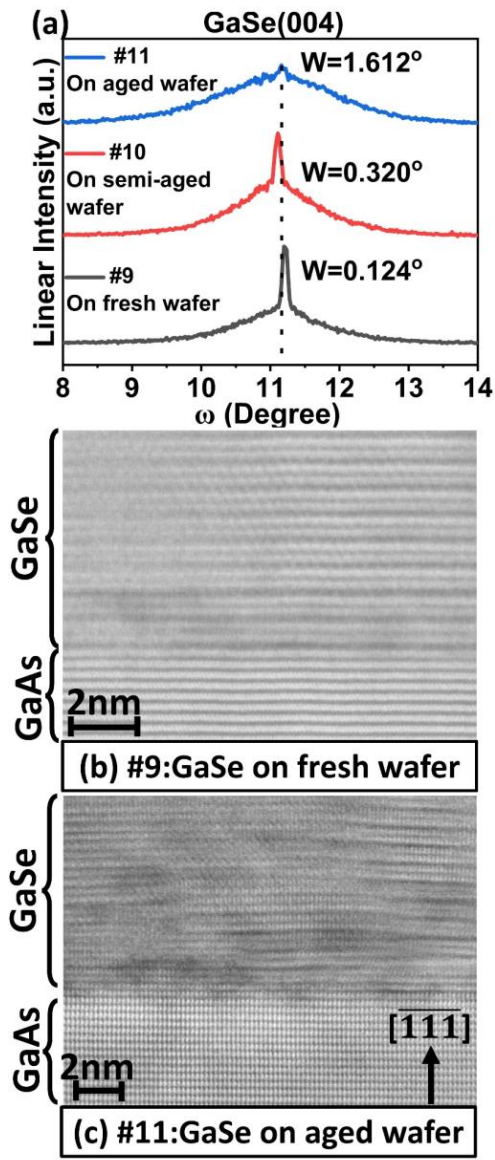

FIG. 9. (a) ω scans of GaSe Sample #9 – #11 around the GaSe (004) plane. "W" indicates FWHM value. Cross-sectional ADF-STEM images (low-pass filtered to reduce noise) of GaSe Sample (b) #9 and (c) #11 grown on a fresh and an aged substrate, respectively. Both samples were grown under the same conditions.

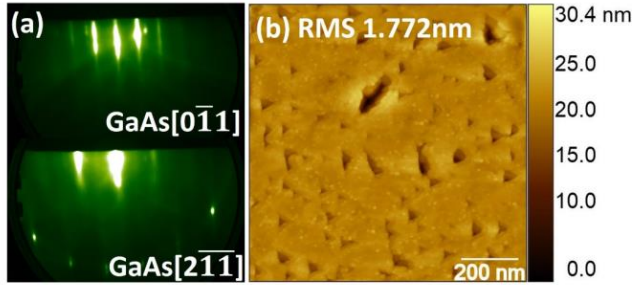

FIG. 10. (a) RHEED pattern and (b) AFM image of a GaAs(111)B wafer that was first exposed to UV-ozone for 5 min and then was thermally deoxidized under a Se flux. The annealing temperature was 680 ºC, the Se flux was $1 \times 10^{14}$ atoms cm$^{-2}$ s$^{-1}$, and the annealing time was 14 min, at which point the sharp streaky RHEED pattern was just visible.

## V. SUMMARY AND CONCLUSIONS

In conclusion, we investigate the thermal deoxidation of GaAs(111)B substrates under a Se overpressure in the ultra-high vacuum environment of MBE and provide the optimal parameters: Se flux $\geq 1 \times 10^{14}$ atoms cm$^{-2}$ s$^{-1}$, annealing temperature of 680 ºC, and annealing time of 7 min. Using this approach, we achieve deoxidation and Se-passivation of GaAs(111)B simultaneously and obtain a smooth platform for subsequent vdW chalcogenide film growth. This approach is highly reproducible. Furthermore, we demonstrate the success and importance of Se-passivation by comparing the GaSe growth on the As-deoxidized and Se-deoxidized substrates, respectively. Theoretical calculations illustrated that the surface configuration is the most energetically favorable when Se atoms replace 75 % of the surface As atoms, therefore, it is likely the surface obtained by this optimal deoxidation/Se-passivation method. We also found that the surface hydrophobicity of GaAs(111)B increases as the substrate ages, possibly caused by a denser oxide layer. Even after deoxidation/Se-passivation using optimal conditions, aged substrates can still affect subsequent sample growth, as removing more oxides appears to cause more damage to the surface. UV-ozone treatment can restore surface hydrophilicity of GaAs(111)B and cause the thickening of the oxide layer and the birth of a new oxide, $As_2O_5$. Finally, storing GaAs(111)B in food-grade vacuum packaging bags can only delay aging, but cannot completely prevent it. Storage in $N_2$-purged containers may be more effective in preserving wafer condition than using food-grade vacuum sealed bags. This work offers valuable insights and experience on the preservation and treatment of GaAs(111)B substrate for growing epitaxial vdW chalcogenide films, thereby developing its applications in heterojunction devices, epitaxial growth and other fields.

## ACKNOWLEDGMENTS

*This study is based upon research conducted at the Pennsylvania State University Two-Dimensional Crystal Consortium – Materials Innovation Platform which is supported by NSF cooperative agreements DMR-2039351. M. Y. and S. L. acknowledge funding from the Coherent/II-VI Foundation. S. A. I., J. L. T., and D. R. H. acknowledge generous support through startup funds from the Penn State Eberly College of Science, Department of Chemistry, College of Earth and Mineral Sciences, Department of Materials Science and Engineering, and Materials Research Institute. J. W. and S. B. S. acknowledge funding from the Basic Office of Science of the Department of Energy under Award DE-SC0018025. The authors acknowledge the use of the Penn State Materials Characterization Lab and appreciate Dr. Jeff Shallenberger for XPS measurements and analysis. They also acknowledge the use of the computational facilities associated with the Institute for Computational and Data Science at Penn State University.*

## AUTHOR DECLARATIONS

**Conflicts of Interest**

The authors have no conflicts to disclose.

## DATA AVAILABILITY

Data of this study are available: https://m4-2dcc.vmhost.psu.edu/list/data/ioDSdyi9qaIR